# Quasi-waveguide amplifiers based on bulk laser gain media in Herriott-type multipass cells.

## Johann Gabriel Meyer,* Andrea Zablah, and Oleg Pronin

*Helmut Schmidt Univerisy, Holstenhofweg 85, 22043 Hamburg, Germany*
[*johann.meyer@hsu-hh.de](*johann.meyer@hsu-hh.de)

**Abstract:** We present here a new geometry for laser amplifiers based on bulk gain media. The overlapped seed and pump beams are repetitively refocused into the gain medium with a Herriott-type multipass cell. Similar to a waveguide, this configuration allows for a confined propagation inside the gain medium over much longer lengths than in ordinary single pass bulk amplifiers. Inside the gain medium, the foci appear at separate locations. A proof-of-principle demonstration with Ti:sapphire indicates that this could lead to higher amplification due to a distribution of the thermal load.

## 1. Introduction

Solid-state laser systems are often based on master oscillator power amplifiers (MOPA), where a laser oscillator provides radiation that is optimized for certain output parameters without being limited by power considerations. The desired power level is then reached by the amplification of the oscillator output in sometimes several successive stages of laser power amplifiers. Each of the stages needs to be designed to work with certain input parameters. With increasing power levels, the management of thermal and nonlinear effects becomes a crucial consideration. In optically pumped solid-state laser gain media, heat is generated due to the quantum defect between laser and pump wavelengths and the presence of non-radiative processes. Excellent thermal conductivity for efficient heat extraction is typically offered by amplifier gain media based on crystalline host materials. Nevertheless, when these materials are applied in simple bulk geometries, thermal gradients will eventually lead to lensing and beam distortions [1]. Furthermore, high temperatures will increase the probability of non-radiative loss processes [2]. Finally, these effects will limit the attainable output power at diffraction limited beam quality. Alternatively, amplifiers based on fiber gain media provide a geometry for efficient heat removal [3,4]. The gain medium is stretched along the propagation direction. The heat load is thereby stretched over a distance several times longer than bulk laser gain media, which reduces the specific heat load per volume. Additionally, cooling becomes more efficient due to the much larger surface of the fiber, even though the thermal conductivity of the glass-made fibers is typically much lower than that of crystalline media. Diffraction limited beam operation and partial suppression of thermal lensing is ensured due to the confinement of the laser mode by the wave-guiding nature of single mode fibers. However, direct amplification of high peak power laser sources in fiber amplifiers is rather limited due to high peak intensities generated inside the fiber. In this case, amplifiers based on the thin-disk [5], or the innoslab geometry are beneficial [6,7].

There are already approaches which could combine certain advantages of bulk and fiber based laser gain media. For standard bulk media, the area of power extraction is limited to a small region around a tight focus inside the bulk medium, which extends typically several mm. Naturally, the main heat load is located there as well, which makes heat extraction difficult and leads to an early onset of power limiting thermal effects. The heat load can be distributed when cascading several gain elements consecutively. This approach has been discussed in the context of periodic resonators [8,9]. Considering only basic thermal lensing, the approach predicts linear power scaling with the number of gain elements. In another approach a rod-like gain medium is extended in length to several cm and reduced in diameter to about a millimeter [10–12]. These so called single-crystal fibers offer advantages of improved heat extraction due to the increased cooling surface similar to fiber gain media, whereas the medium itself can be made of crystalline material and offers therefore high thermal conductivity. Though, in contrast to fibers, these media offer no wave guiding due to their still relatively large diameters. Therefore, the amplified seed beam is focused only to a shallow caustic inside the gain medium, which will increase the threshold for efficient power extraction due to the reduced intensity.

We propose here an amplifier concept, which relies on the repetitive refocusing or guiding of the laser radiation inside a bulk laser gain medium placed inside a Herriott-type multipass cell (see Fig. 1). The imaging property of the cell allows for the collinear propagation of amplified seed and pump beam, leading to a good overlap of both beams. (In contrast to multipass amplifiers, the seed beam is not sent back into the same region of the gain medium. [13]) Refocusing inside the cell happens at spatially separated foci (see Fig. 1), which effectively increases the propagation length inside the gain medium, while the refocusing ensures that inside the gain medium the beam is small enough for efficient power extraction. Due to its quasi-waveguide structure, this amplifier design shares similarities with core pumped fiber amplifiers, although it can be applied to many kind of bulk laser gain media. While extending the propagation distance by repetitive focusing into the gain medium, the total heat load inside the gain medium is distributed over the consecutive passes. Thereby the temperature can be reduced for each individual pass through the gain medium. For some laser gain media, e.g. Ti:sapphire, Cr:ZnS/Se, the absolute temperature inside the laser gain medium can have through quenching effects a significant impact on the fluorescence lifetime of the upper laser level and thereby limit the performance at higher output powers [2,14]. Reducing the temperature inside the gain medium would be therefore beneficial for high power laser amplification in these gain media.

For some bulk laser gain media, high doping concentrations of active laser ions can enhance quenching processes [1,15–17]. These quenching processes will lead to a loss of excited active laser ions and thereby limit the amplification performance of the laser gain medium. The lower the acceptable doping concentration, the longer the propagation length required to absorb a major part of the pump radiation. Thus a correspondingly long laser gain medium is required, over which the pump radiation is absorbed. In the case of longitudinal pumping, the seed beam is propagating collinear with the pump beam. Tight focusing of the seed beam, which is often necessary for efficient power extraction, might produce a sufficiently small seed beam diameter over a distance much shorter than the length of the gain medium, i.e. the propagation length of the pump radiation. Therefore, efficient power extraction might be constrained to only a fraction of the pumped region of the laser gain medium. Here, it could be beneficial to employ the refocusing property of a Herriott-type multipass cell. When the gain medium is place at the focal plane of the cell, both pump and seed radiation are passing together several times through the gain medium. Thereby the propagation length is increased and the pump power can be absorbed over a longer distance, while the seed beam diameter stays small enough for sufficient power extraction over the whole absorption path length of the pump radiation.

## 2.  Theoretical considerations

Like with standard bulk amplifiers, there are principally two ways how to optically pump the proposed quasi-waveguide amplifier based on a Herriott-cell. The pump radiation can be propagated collinearly with the seed beam, which corresponds to longitudinal pumping in standard bulk amplifiers, or the pump radiation can be introduced from another direction, which can be beneficial for pumping by using laser diodes.

For collinear pumping, the seed and the pump beam are overlapped and coupled together into the Herriott-cell (see Fig. 1). This pumping scheme allows for high extraction efficiency as it can provide the highest overlap between pump and seed beam. Consequently, it is crucial that not only the seed, but also the pump beam have close to diffraction limited beam quality. Otherwise, the beam radius of the pump beam inside the cell would be substantially larger than the radius of the seed beam. This could lead to an unsatisfactory overlap between pump and seed beam, as well as to clipping, and therefore to decreased performance of the amplifier. Furthermore, to achieve reproducible focusing at the position of the gain medium after each pass in the cell, it is also necessary to match the beam caustic of seed and pump beam to the corresponding eigenmodes of the Herriott-cell. The waist radius of the eigenmode of a symmetric Herriott-cell is given by [18,19]:

$$w_{0,eigen} = \sqrt{\frac{\lambda_0}{2\pi} R \sin\left(\pi \frac{M}{N}\right)} \qquad (1)$$

It is dependent on the vacuum wavelength $\lambda_0$, the radius of curvature of both cell mirrors $R$, the number of reflections on one cell mirror $N$, and the cell configuration parameter $M$ $(1, 2, ..., N-1)$, which determines where after one reflection the next reflection is appearing in relation to the spot pattern formed by all $N$ reflections on one mirror. The waist radius of the eigenmode determines which intensity can be reached inside the gain medium. Generally, the seed beam intensity has to be sufficiently high to lift the rate of stimulated emission inside the gain medium well above the rate of spontaneous emission and thereby enable efficient power extraction from the amplifier. To reach a high efficiency, the cell parameters have to be chosen corresponding to their relation to the waist radius. Depending on the waist radius, the beam diameter is quickly expanding in normal bulk media due to diffraction. To limit the seed beam diameter inside the gain medium, its thickness might be restricted to about the Rayleigh range of the cell eigenmode.

The number of reflections on each mirror is limited by the size of the mirrors and the size of the beams. The mode radius on the cell mirrors is given by [18,19]:

$$w_{m,eigen} = \frac{w_{0,eigen}}{\cos\left(\frac{\pi}{2}\frac{M}{N}\right)} = \sqrt{\frac{\lambda_0}{2\pi} R \sin\left(\pi \frac{M}{N}\right)} \sec\left(\frac{\pi}{2}\frac{M}{N}\right) \qquad (2)$$

It becomes clear that over the whole caustic the beam radius scales with the square root of the wavelength. It is therefore desirable to pump the laser gain medium at a wavelength close to the seed wavelength to ensure a good overlap inside the gain medium.

The collinear pumping is not only providing high overlap; it also reduces the demand for high laser ion doping concentrations. It might even become desirable to absorb the pump radiation only after multiple passes through the laser gain medium. Gain media which suffer from quenching due to high doping concentration could therefore benefit from this pumping scheme.

### 3. Experimental setup

A proof of concept experiment for the Herriott-cell quasi-waveguide amplifier design described above was developed. The purpose of this experiment was to investigate whether the amplifier could be experimentally realized, whether amplification could be measured, and whether this design had any advantages over the standard single pass amplifier.

Ti:sapphire was chosen as the gain medium in this experiment for several reasons. Ti:sapphire has excellent thermal conductivity. This easily available bulk medium also has broad absorption and emission cross sections, allowing the use of existing diffraction limited lasers as pump sources, essential for a good beam overlap and extraction efficiency. The pump sources for Ti:sapphire are found within the visible wavelength range, which facilitates the alignment process. Alignment is particularly crucial here for getting optimal gain results with the multipass cell, as the pump and seed beam need to overlap for all the multiple passes across the Ti:sapphire crystal. The experimental demonstration was fostered by the availability of the laser sources with the corresponding wavelengths at our research institution and the commercial availability of further customized amplifier components for Ti:sapphire. The gain medium was pumped with a Verdi V10 CW 532 nm laser *(Coherent)*, with $M^2 < 1.1$ and 10 W of output power at the setup. The seed was a CW Ti:sapphire laser *(M Squared Lasers)*, with $M^2 < 1.1$, tuned at 800 nm and delivering 1.4 W of output power at the setup.

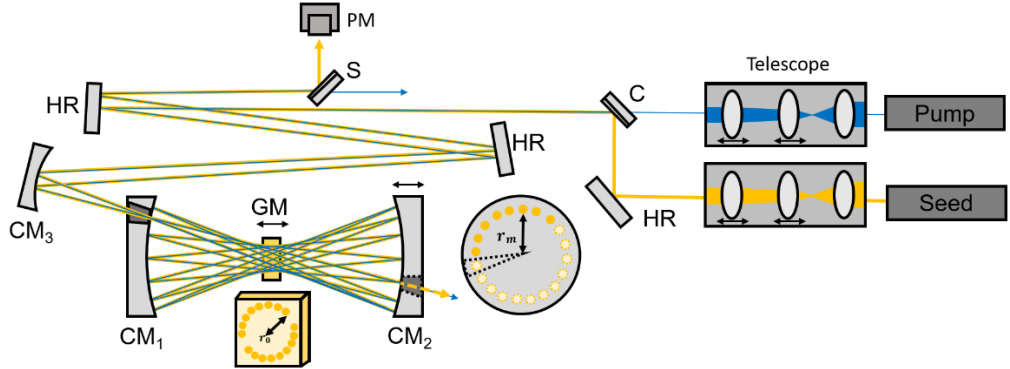

*Figure 1: Schematic of the experimental setup for the quasi-waveguide amplifier based on a Herriott-type multipass cell with a bulk laser gain medium (GM) and collinear pumping. The pump and seed beams were mode matched by two telescopes having each two UVFS plano-convex lenses with f = 100 mm, and one UVFS aspheric lens at the center with f = 50 mm. They were then overlapped by a dielectric 45° beam combiner (C) transmitting the pump and reflecting the seed beam. The beams were focused with a curved HR dielectric mirror with a ROC of 300 mm ($CM_3$) into the multipass cell. The symmetric Herriott-cell consisted of two HR (R = 99.8 %) dielectric mirrors with a ROC of 50 mm ($CM_1$ and $CM_2$). $CM_1$ had a cut at the side for the beams entering and leaving the Herriott cell. $CM_2$ shows with dashed lines the possibility of making a second cut to couple out the beams early to avoid additional reflectivity losses after most of the pump has been absorbed. The beams are repetitively refocused into the 5 x 5 mm Ti:sapphire crystal of 2 mm length and 0.8 $cm^{-1}$ absorption coefficient. Lastly, the amplified seed beam was isolated using a 45° separator mirror (S), and sent to a power meter (PM). The Herriott cell pattern on the mirror and crystal are also shown, with their corresponding radii, $r_m$ and $r_0$ respectively (lighter colored spots represent the missing reflections when coupling out early).*

The pump and the seed beam were collinearly propagated inside the Herriott cell. To repetitively refocus the radiation into the gain medium with the same waist diameter, it was necessary to match the modes of the pump and the seed beam to the corresponding eigenmodes of the cell. As described above, for a good beam overlap, the wavelengths of the seed and the pump would ideally be closely matched. The eigenmode radius scales proportional to the square root of the wavelengths (see Eq. 1). For the setup presented here, using an 800 nm seed beam pumped by a 532 nm laser, the mode radius of the seed was everywhere in the Herriott-cell estimated to be 1.2 times larger than that of the pump.

In order to match the pump and seed beams' caustic with the corresponding eigenmodes of the Herriott-cell, two separate three-lenses telescopes were used, one for each beam (see Fig. 1). Both the beam size and divergence angle could be independently adjusted. Mounting the lenses on translation stages allowed for fine adjustments to increase the extraction efficiency.

To reduce losses to spontaneous emission in the gain medium, the seed beam needs to exceed the saturation intensity. For Ti:sapphire the saturation intensity is 200 kW/cm². As the power of the seed laser was limited to 1.4 W, the intensity was controlled by the beam radius inside the gain medium, which is directly linked to the configuration of the Herriott cell (Equation 1). To achieve a small waist radius, a rather small radius of curvature of 50 mm was chosen for both cell mirrors. Furthermore, the waist radius could be reduced by aligning the Herriott cell close to a concentric resonator configuration, where the distance of the cell mirrors approaches the sum of the radii of curvature of the two cell mirrors. For a Herriott cell, this is realized by bringing the ratio of the configuration parameter $M$ and the number of reflections per mirror $N$ close to one. Consequently, the longest cell configuration results for $M = N-1$. Then, the cell length increases and the waist radius decreases monotonically with the number $N$. On the cell mirrors with a diameter of 25 mm it was possible to fit $N = 20$ reflections without clipping on the cut in the mirror. The resulting seed waist radius was $w_0(800\text{ nm}) = 32$ µm and pump waist radius $w_0(532\text{ nm}) = 26$ µm. Considering the current setup, with the calculated waist radius and the available seed power, it was possible to reach a peak intensity inside the gain medium of just 90 kW/cm², which is about a factor two below the saturation intensity.

The employed gain medium had a length of 2 mm and was manufactured with a c-cut *(Roditi)*. A low absorption coefficient of 0.8 cm⁻¹ at 532nm (FOM > 200) was chosen so that the pump radiation could be passed several times through the gain medium before complete absorption. Most of the pump radiation, 93 %, was calculated to be absorbed after 17 passes through the gain medium. However, the cell configuration (N = 20) chosen to optimize power extraction would lead to 40 passes through the gain medium, exceeding the number necessary for sufficient pump absorption. For further passes it is expected that the introduced losses on the coatings will surpass the additional gain. Hence, instead of passing the beams completely through the Herriott cell, the beams were coupled out early through a cut in the second cell mirror, resulting in only half a circle of eight reflection spots on each cell mirror (see Fig. 1). It is noteworthy that even when coupling the beams out after only 17 from 40 possible passes in the Herriott cell, the beams passing the gain medium were lying on an almost complete circle, given the opposing reflections from each mirror for the longest configuration (M = N - 1). The radius of the beam pattern inside the gain medium $r_0$ (Fig. 1) was calculated from the pattern radius at the cell mirrors $r_m = 9.5$ mm and the cell configuration parameters M and N, according to the following equation:

$$r_0 = r_m \cos\left(\frac{\pi}{2}\frac{M}{N}\right) \qquad (3)$$

The resulting radius of $r_0 = 0.75$ mm fitted well within the gain medium with an aperture of 5 mm x 5 mm. The gain medium was mounted in a water cooled aluminum heat sink. The temperature of the cooling water was 20 °C. Indium foil with a thickness of 0.1 mm was used to improve the thermal contact. Thermal lensing effects could be corrected to some extent by adjusting the cell's length and crystal position at focus with translation stages.

### *Single pass amplifier as reference*

The Herriott cell amplifier was compared to a single pass amplifier with similar parameters. Instead of sending the overlapped pump and seed beams into the Herriott cell, they were directly focused (f = 400 mm) into another highly doped Ti:sapphire crystal. For an equal comparison, the pump and the seed beam were set to match the Herriott cell's waist radius using the three-lens telescopes. The single pass amplifier allowed the advantage of having both beams perfectly overlap under the same focus waist, set here to $w_0 = 32$ µm, which was carefully aligned with a beam profiler at the focus position. To account for having one single beam path across the gain medium, a crystal with higher absorption coefficient of 7 cm⁻¹ at 532 nm (FOM > 200) was needed, manufactured with an a-cut. To absorb the same amount of pump radiation of

93 %, as in the Herriott-cell amplifier, during on single pass, the length of the gain medium had to be extended to 4 mm considering that no higher absorption coefficients were available. The lateral dimensions of the gain medium were maintained at 5 x 5 mm. The gain medium was mounted on an identical heat sink. In comparison to the Herriott cell amplifier, it is noted that the distance from the beam position, i.e., the heat source to the heat sink is increased, as the beam pattern formed by the Herriott cell brings the beams closer to the edge of the gain medium, if the lateral dimensions are maintained. This could increase the temperature in the gain medium additionally to the local concentration of all deposited heat. In further investigations, it would be possible to reduce the size of the gain medium for the single pass amplifier to match the distances from the beams to the heat sink.

### *Power characterization procedure*

To compare the resulting gain from both amplifiers under equivalent conditions, the amplified seed power was recorded using the same procedure. The seed power was kept constant during the measurements. The pump power was increased in steps of 1 W, from 1 W output power up to 10 W. First, the seed beam was turned on and its power was recorded (*Ophir* high sensitivity thermal sensor) after passing it through the unpumped amplifier and being reflected by the separator. Then, the pump laser was turned on and the amplified seed power was recorded. The gain was measured as the ratio between the power extracted from the amplifier and the power of the unamplified seed beam. Lastly, the pump power was turned off, and the crystal was left to cool down to room temperature before repeating the procedure for the next pump power level.

To discard the possibility of measuring parasitic pump reflections rather than amplified seed power, the pump reflected from the separator was measured for both amplifier setups in the absence of the seed beam. For the Herriott-cell amplifier, no parasitic pump reflections could be recorded with the sensitivity of the measurement device. For the single pass amplifier, a small parasitic pump reflection could be measured at the output. This value was subsequently subtracted from the amplified seed power for each measurement.

For higher pump powers, the temperature of the gain medium increased and thermal lensing effects became noticeable. The output beam deformed into a reniform shape and the amplified power decreased. The thermal effects could be partially dealt with, to the extent of the setup's capabilities, by adjusting the cell distance and crystal position, and in the case of the single pass amplifier, only the crystal position. After performing this correction during measurements, it was possible to recover the output beam's round shape and to increase the amplified seed power.

A thermal imaging camera (*Testo*) was used to monitor the temperature of the gain medium during pumping. The temperatures were corrected by the emissivity of the Ti:sapphire crystal. The emissivity was measured to be $\varepsilon = 0.5$ at 100 °C, when heating the Ti:sapphire crystal on a temperature controlled hot plate. The recorded crystal temperature during amplification was corrected with the measured emissivity.

## 4. Results

A characterization of the power handling capabilities of each amplifier, Herriott-cell and single pass, was carried out by comparing their gain.

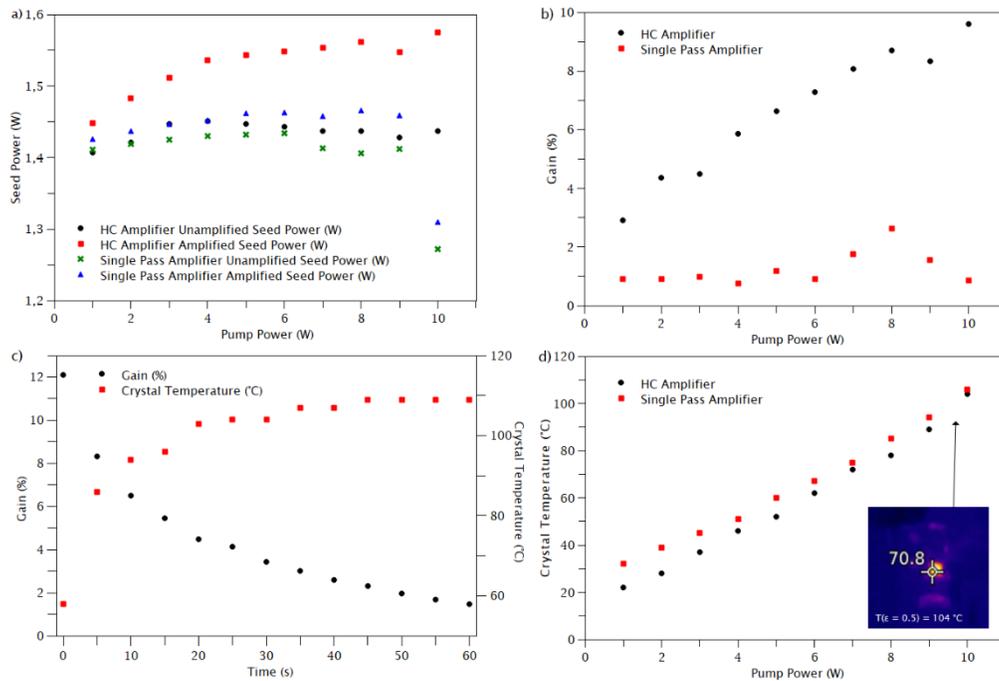

*Figure 2: **a)** Measured output power characteristic of unamplified and amplified seed beams of the Herriott-cell and single pass amplifiers for pump powers from 1 to 10 W. **b)** Measured gain for the Herriott-cell (black) and single pass (red) amplifiers. **c)** Development (decay) of the gain (black) due to high crystal temperature (red) during the first minute of pumping at full power of 10 W. **d)** Measured crystal temperature for the Herriott-cell (black) and single pass (red) amplifiers from 1 W to 10 W pumping power. The inset shows a thermographic picture of the gain medium of the Herriott cell amplifier with an emissivity corrected temperature of 104 °C at a pump power of 10 W pumping. The thermal image looks similar for both Herriott-cell and single pass amplifiers. With the current camera resolution, only an average temperature could be measured, as the camera was not capable of resolving the individual location of each beam in the gain medium.*

### *Herriott-cell amplifier*

The power characterization from 1 – 10 W of pump power in 1 W steps was carried out as described above. For this process, the amplifier was optimized once, during the first amplification measurement at 1 W pump power. The remaining measurements were carried out consecutively. After turning off the pump beam at the end of each measurement with increasing pump radiation, small power variations could be detected on the unamplified seed power due to remaining thermal effects at the crystal, i.e. the output power was not constant. The recorded gain and crystal temperature increased monotonically while increasing the pump power up to 10 W (Fig. 2b). The gain increased up to a value of almost 10 % at the highest pump power. The maximum temperature reached by the Ti:sapphire crystal during amplification at 10 W pump power was 109 °C (Fig. 2d). A waiting time of 3 min was necessary to cool down the crystal back to room temperature after amplifying at full pump power.

From the power characterization measurements, it was noticed that at crystal temperatures of 46 °C and higher, corresponding to pump powers of 4 W and above (Fig. 2d), the amplified output power quickly decayed for a constant pump power. At lower pump powers, the amplified seed power remained stable. To further investigate this, amplification at 10 W pump power was once again measured. Starting at an optimized gain of 12 %, the amplifier was pumped for one minute. Within this period the gain dropped to 2 % (Fig. 2c). After pumping the gain medium for 10 min, the gain completely disappeared.

Compared to the initial power characteristic measurement it was possible to increase the maximum gain slightly with an improved alignment procedure. Careful optimization was done at low pump power (3 W) before directly increasing the pump to full power (10 W). Thermal lensing effects were first corrected. The pump and seed beam overlap, and waist size at the gain medium were then fine-tuned for maximum output power. Then, the pump power was increased to 10 W. Immediately after that, a gain of 9 to 13 % could be measured during several repetitions of the procedure. During this procedure, the maximum gain could not be further optimized during pumping due to the quick decay in amplified power at high temperatures of the crystal. After attempting optimization, thermal effects had already depleted most of the gain, and only a fraction of the amplified power could be recovered.

### *Single pass amplifier*

The same power characterization procedure as for the Herriott-cell amplifier was followed for the single pass amplifier. In this case, the ratio between the seed beam during pumping and in the absence of pumping was on the order of one percent and it did not exceed 2.6 % (Figure 2a and b). Even though the amplification was small, it clearly disappeared when the pump beam was blocked.

The crystal temperature increased from room temperature at low pump power up to 106 °C at full pump power, similar as with the Herriott-cell amplifier (Fig. 2d).

The gain for the single pass amplifier could be slightly optimized with the same procedure as for the Herriott-cell amplifier, where the setup is carefully optimized once at 3 W pump power, and then immediately increasing the pump to 10 W. The highest measured gain was 4 %, three times lower than the recorded Herriott-cell amplifier gain. A power decay of the amplified beam over time was also seen here when the amplifier was pumped at full power. When the procedure was repeated several times, gain values between 1 – 4 % were usually recorded.

### *Amplifier comparison*

**Table 1. Herriot cell and single pass amplifier experimental conditions and results**

| Amplifier | Gain medium | | | | | | |
|---|---|---|---|---|---|---|---|
| | Material | Length | Aperture | Cut | α | FOM | Propagation length |
| Herriott cell | Ti:sapphire | 2 mm | 5 x 5 mm$^2$ | c-cut | 0.8 cm$^{-1}$ | > 200 | 34 mm |
| Single pass | | 4 mm | | a-cut | 7 cm$^{-1}$ | | 4 mm |

| Amplifier | Seed (800 nm) | | Pump (532 nm) | | Output at 10 W pump | |
|---|---|---|---|---|---|---|
| | $w_0$ | Power | $w_0$ | Absorption[1] | T[2] | Max. gain[3] |
| Herriott cell | 32 μm | 1.4 W | 26 μm | 93 % | 109 °C | 13 % |
| Single pass | | | 32 μm | 93 % | 106 °C | 4 % |

[1] Calculated pump absorption based on the gain medium's absorption coefficient and beam's path length.
[2] Corrected with emissivity of $\varepsilon = 0.5$
[3] Mostly decayed after 1 minute.

As seen on Figure 2 (a) and (b), the measured amplification was at least three times higher for the Herriott cell amplifier with a 13 % gain than for the single pass amplifier with a maximum of 4 % gain. The amplification with the Herriott cell monotonically increased with increasing pump power, while it stayed almost constant for the single pass amplifier. Thus, for achieving the highest possible gain under the same experimental conditions, the Herriott cell design proved beneficial. However, with the Herriott cell amplifier it was not possible to keep the amplification at pump powers higher than 4 W, since high temperatures, already above 46 °C, led to a decay in the output power.

Figure 2 (d) shows similar crystal temperatures when incrementing the pump power on both amplifiers. While about the same pump power was absorbed by both gain media, the temperature increased similarly in the two cases, from room temperature when unpumped up to 110 °C when pumping at 10 W. Due to the low camera resolution (Fig. 2d), the camera could not resolve the individual location of the beams and it showed only an average temperature. As the Herriott-cell amplifier delivered significantly more output power than the single pass amplifier, it is assumed that locally at the seed beams of the Herriott-cell amplifier, but unresolved by the camera, the temperature was lower. Therefore, the power limiting temperature effects would have played a smaller role compared to the single pass amplifier.

## 5. Conclusion

In summary, we presented an amplifier concept resembling a quasi-waveguide approach for the use with bulk laser gain media, in this case realized with Ti:sapphire. The amplified seed radiation is refocused into several distinct spots at the solid-state laser gain medium by means of a Herriott-type multipass cell. The pump radiation was coupled collinearly with the amplified seed radiation into the cell. Similar to fiber amplifiers, the beam diameters inside the gain medium stayed here confined over a much longer propagation length of 34 mm for the Herriott cell. In comparison, amplifiers with bulk laser gain media usually have only a short propagation length of e.g. 4 mm for the single pass amplifier presented here, or 14 mm for a hypothetical single pass amplifier design assuming a crystal length similar to the beam's Rayleigh range. The heat load brought into the gain medium was here distributed over a larger volume, across several spots at the crystal. As a result, thermal effects were less detrimental and the gain extraction was more efficient. This was demonstrated by comparing the Herriott-cell amplifier with a single pass amplifier using analogous parameters. Even though the single pass setup had some advantages over the Herriott cell, such as having twice the crystal length which should have improved the cooling of the gain medium, and having a better overlap of the pump and seed beam with the same waist size at the crystal, it still performed poorly in comparison. The Herriott cell amplifier yielded a gain of at least four times higher, up to a maximum of 13%. Thus, having multiple beam passes distributed across the gain medium resulted in better gain extraction for the Herriott-cell amplifier. However, the seed power on both amplifier designs was adversely affected when exposed to high temperatures at the gain medium due to high pumping radiation, as seen by the quick decay in amplified power. Another limitation for the Herriott-cell amplifier was its incapacity to reach the gain medium's saturation intensity. Higher power seed lasers can be used to achieve higher intensities. Alternatively, a smaller Herriott cell using shorter radius of curvature mirrors, if available, could be built. It would then be possible to reach the ideal waist size to attain the required saturation intensity with the multipass cell for a good extraction efficiency. This, nonetheless, would also introduce additional technical and alignment challenges. Other gain media with a lower saturation intensity would require lower powers to efficiently extract from the amplifier. The gain medium Cr:ZnS is of relevance for the generation of powerful coherent mid-IR sources [20,21]. It has a saturation intensity of merely 15 kW/cm$^2$. As the mode area scales with the wavelength, the determining factor is rather the σ-τ product, which is still more than four times higher for Cr:ZnS compared to Ti:sapphire. Correspondingly, the power would only need to be four times lower for Cr:ZnS to reach the saturation intensity. A larger mode radius on the cell mirrors due to the longer wavelength could make it necessary to reduce the number of reflection. This could increase the waist radius, which reduces the peak intensity. However, for the presented configuration, it would not necessarily play a limiting role.

**Funding.** The research was completely financed by our university budget. No external funding.

**Acknowledgments.** We would like to thank Kilian R. Fritsch, formerly affiliated with Helmut Schmidt University, for the discussions and advice on the topic of Herriott cells.

**Disclosures.** The authors declare no conflicts of interest.

**Data availability.** Data underlying the results presented in this paper are available on request.